\documentclass{article}
\usepackage{spconf,amsmath,graphicx,hyperref,multirow,multicol,booktabs,siunitx,fontsize}
\usepackage[table]{xcolor}
\usepackage[protrusion=false,expansion=true]{microtype}

\newcommand{\yedit}[1]{\textcolor{black}{#1}} 
\newcommand{\sedit}[1]{\textcolor{black}{#1}}
\newcommand{\ssedit}[1]{\textcolor{black}{#1}}
\newcommand{\sssedit}[1]{\textcolor{black}{#1}}

\title{CC-G2PnP: Streaming Grapheme-to-Phoneme and prosody \\ with Conformer-CTC for unsegmented languages}
\name{Yuma Shirahata, Ryuichi Yamamoto}
\address{LY Corporation}
\begin{document}
\changefontsize[11.1pt]{9.1pt}
%
\maketitle
\begin{abstract} 
We propose CC-G2PnP, a streaming grapheme-to-phoneme and prosody (G2PnP) model to connect large language model and text-to-speech in a streaming manner.
CC-G2PnP is based on Conformer-CTC architecture. Specifically, the input grapheme tokens are processed chunk by chunk, which enables streaming inference of phonemic and prosodic (PnP) labels. By guaranteeing minimal look-ahead size to each input token, the proposed model can consider future context in each token, which leads to stable PnP label prediction. 
Unlike previous streaming methods that depend on explicit word boundaries, the CTC decoder in CC-G2PnP effectively learns the alignment between graphemes and phonemes during training, making it applicable to unsegmented languages. Experiments on a Japanese dataset, \ssedit{which has no explicit word boundaries}, 
show that CC-G2PnP significantly outperforms the baseline streaming G2PnP model in the accuracy of PnP label prediction.

\end{abstract}
\vspace{-1mm}
\begin{keywords}
Grapheme-to-phoneme, Conformer-CTC, streaming, prosodic label prediction
\end{keywords}

\section{Introduction}

Owing to the growing demand for natural and efficient human–machine interaction, research on spoken dialogue models has been actively studied~\cite{ji2024wavchat}. Although one line of research explores end-to-end models that perform speech-to-speech generation within a single model~\cite{defossez2024moshi,xie2024mini,nguyen2023generative}, cascade approaches that combine automatic speech recognition (ASR), a large language model (LLM), and text-to-speech (TTS) offer advantages in robustness and flexibility~\cite{huang2024audiogpt,dubey2024llama}. To improve the response speed of spoken dialogue models, cascade approaches have applied streaming TTS to the text incrementally generated by the LLM, achieving promising results~\cite{dubey2024llama,du2025instantspeech,dekel2024speak,ma2019incremental}. 

\ssedit{In terms of TTS input, some models can take the raw output of the LLM, i.e., graphemes~\cite{du2024cosyvoice,du2024cosyvoice2}. These models are trained to generate correct pronunciation and prosody in speech directly from graphemes. However, achieving sufficient performance in this setting generally requires a huge amount of text-speech paired data, since the model must simultaneously learn pronunciation and prosody, as well as additional aspects of speech such as speaker characteristics and speaking style.}
Therefore, a practical alternative is to first convert graphemes into phonemes and prosodic labels (G2PnP) and then use them as input to the TTS model~\cite{park2022unified,shirahata2024audio,dai2022automatic}.
\sedit{Since phonemes and prosodic labels (e.g., accent position and accent type) provide more direct information to synthesize speech compared to graphemes, this approach tend to yield more stable performance even with limited training data.} 
Nevertheless, since G2PnP is introduced between the LLM and the TTS, reducing latency requires streaming not only in TTS but also in G2PnP.

\ssedit{A naive approach} to achieve streaming G2PnP is to process the input text in chunks using \ssedit{non-streaming} sentence-level G2PnP models~\cite{park2022unified,kurihara2024enhancing}. However, since G2PnP in many languages is highly dependent on the surrounding context, this approach generally struggles to deliver robust performance. 

Since G2PnP is a sequence-to-sequence mapping problem, another approach would be to employ the encoder–decoder Transformer architecture~\cite{vaswani2017attention}. \ssedit{Dekel et al.~\cite{dekel2024speak} proposed LLM2PnP, a streaming G2PnP model that applies} a word-level restriction to the attention mask of the Transformer. \ssedit{Unlike the naive chunk-based streaming approaches}, this model can effectively incorporate both past and future context during streaming, and it demonstrated strong performance in streaming G2PnP for English. 
However, since the implementation of look-ahead and masking assumes the presence of word boundaries, it cannot be directly applied to languages without explicit word segmentation, such as Japanese or Chinese, i.e., unsegmented languages. 

To overcome the limitation of the naive approaches and LLM2PnP, we introduce CC-G2PnP, a streaming G2PnP model capable of incorporating past and future context without dependence on word boundaries. CC-G2PnP is composed of a stack of streaming Conformer~\cite{gulati2020conformer} layers followed by a connectionist temporal classification (CTC) decoder~\cite{graves2006connectionist}. Unlike LLM2PnP, which requires an explicit mapping between words and phonemes \sedit{for making attention masks}, CC-G2PnP learns these mappings dynamically through CTC, making it applicable to unsegmented languages. Furthermore, to enable efficient streaming with the Conformer, the proposed method employs chunk-aware streaming~\cite{noroozi2024stateful}, which divides the input sequence into chunks and allows tokens to attend within each chunk to realize look-ahead. To adapt this framework to the G2PnP task, we further extend the model so that every token has at least one token of look-ahead, thereby enabling more stable PnP label prediction.

To demonstrate the effectiveness of CC-G2PnP in the streaming G2PnP task for unsegmented languages, we conducted experiments using a Japanese dataset. 
The experimental results revealed that CC-G2PnP achieved significant improvements over baseline streaming methods in terms of both objective evaluations (character error rate, CER; sentence error rate, SER) and subjective assessments of the naturalness of TTS samples.

\section{Method}

\label{sec:method}

\begin{figure}[t]
    \centering
    \includegraphics[width=\linewidth]{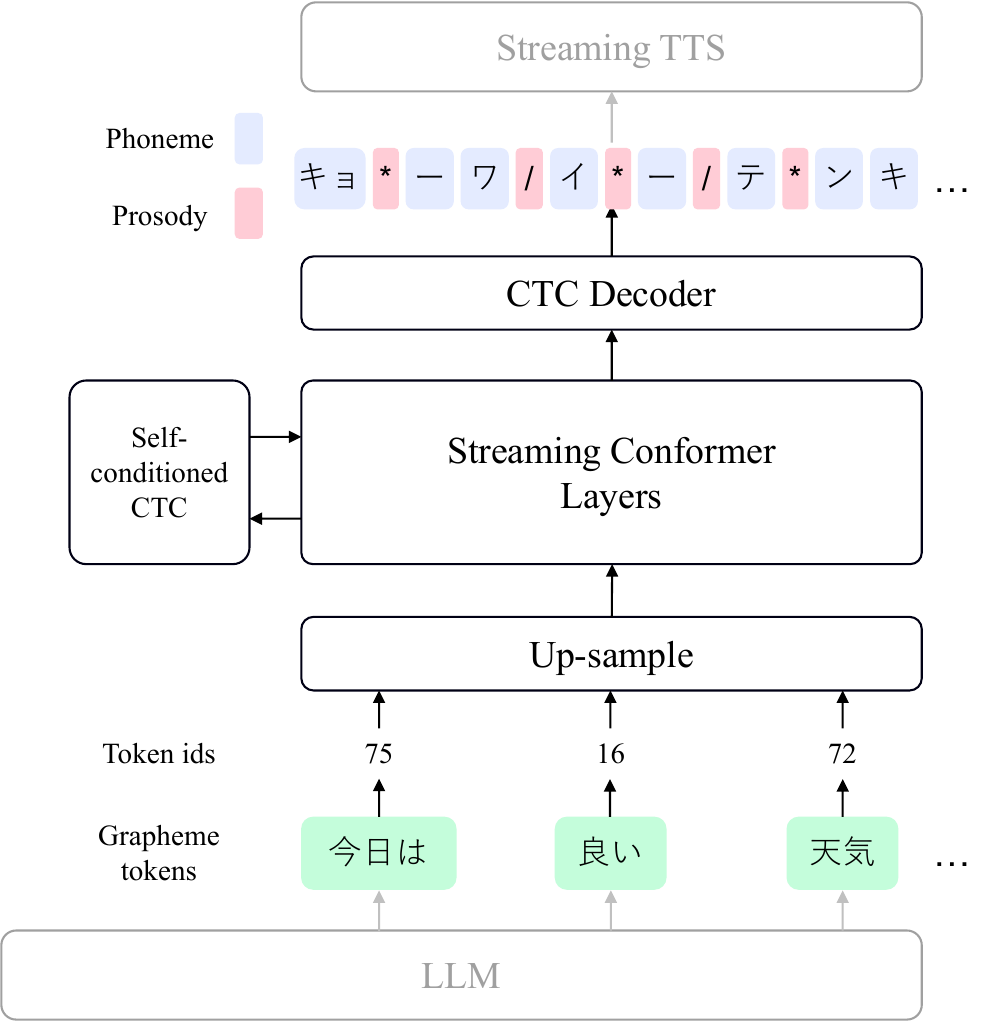}
    \caption{Proposed model architecture. The model takes grapheme tokens as input and predicts a mixed sequence of phoneme and prosodic symbols.}
    \label{fig:model}
\end{figure}

\subsection{Model architecture}
\label{subsec:architecture}
Considering the successful application of CTC in prior work on non-streaming G2P tasks~\cite{rao2015grapheme,wang2023liteg2p,girrbach2023sigmorphon}, as well as its suitability for streaming scenarios~\cite{noroozi2024stateful,moritz2019streaming}, the proposed method employs a CTC decoder to model the \yedit{G2PnP} task as a sequence labeling problem. 
\ssedit{The use of CTC eliminates the need for pre-defined alignments and word boundaries.}
The proposed model architecture is shown in Fig. \ref{fig:model}. As shown in the figure, the proposed model adopts a structure commonly used in the field of ASR, consisting of a stack of Conformer layers~\cite{gulati2020conformer} followed by a CTC decoder~\cite{graves2006connectionist,higuchi2020mask}.  To relax the conditional independence of each token and improve the performance of the model, we introduced self-conditioned CTC~\cite{nozaki2021relaxing} into the intermediate \ssedit{Conformer} layers. 
The model is optimized to minimize the sum of the final and the intermediate CTC losses.

\subsection{Chunk-aware streaming}
To make the Conformer streaming-capable, it is necessary to restrict the dependencies on future tokens in both the convolution layers and the self-attention layers. For the convolution layers, since the kernel size is fixed, future-token dependency can be eliminated simply by using causal convolution. For the self-attention layers, 
while several methods have been proposed to enable streaming~\cite{moritz2019streaming,zhang2020transformer}, in this study we adopt the chunk-aware streaming approach introduced by~\cite{noroozi2024stateful}.  

\begin{figure}[t]
    \centering
    \includegraphics[width=\linewidth]{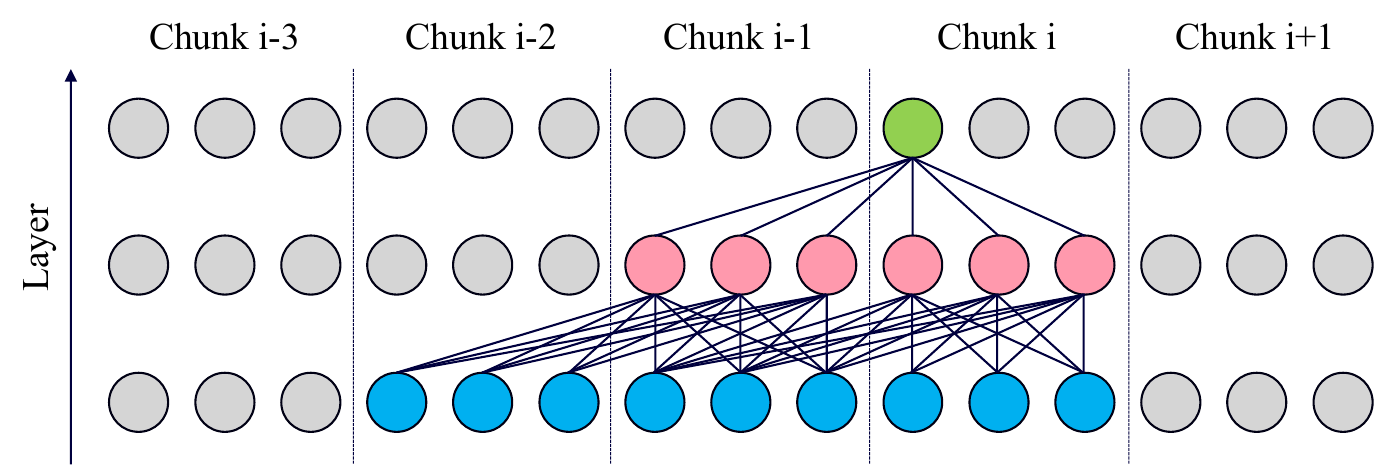}
    \caption{Chunk-aware streaming. The chunk size $C=3$ and the past context size $P=3$. The green token can attend to pink tokens, which correspond to the tokens within its chunk and the past $P$ context.}
    \label{fig:streaming}
\end{figure}

\ssedit{Fig.~\ref{fig:streaming} illustrates the concept of chunk-aware streaming.}
In this approach, the input tokens are divided into chunks of size $C$. 
Each token in a chunk can attend to all other tokens within the same chunk as well as to tokens from a fixed number $P$ of past contexts. 
As shown in the figure, this guarantees look-ahead within each chunk (ranging from $C-1$ to 0), while keeping the number of future tokens referenced constant across self-attention layers. 
This is a major advantage over regular look-ahead, where the number of referenced future tokens grows linearly with the number of layers, resulting in increased latency.

\begin{figure}[t]
    \centering
    \includegraphics[width=\linewidth]{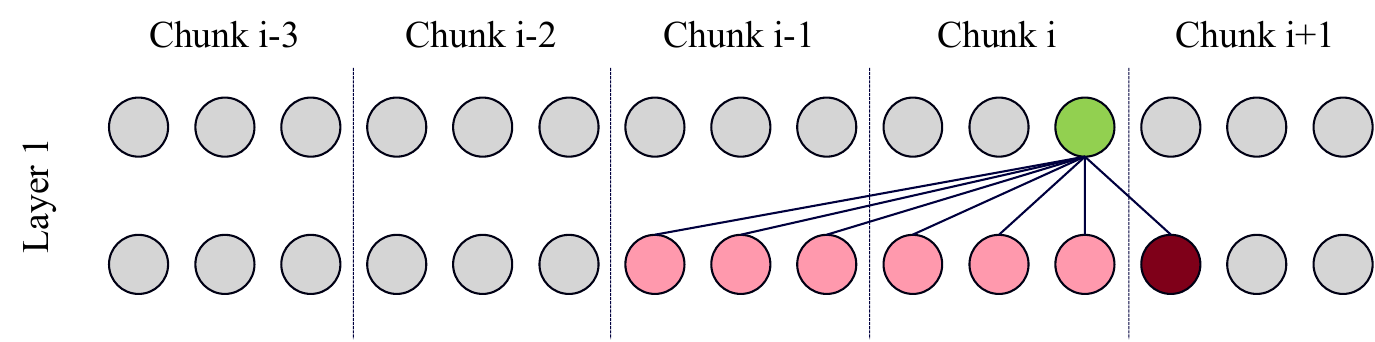}
    \caption{minimum look-ahead (MLA). MLA allows the first layer of self-attention to reference future tokens outside the current chunk, thereby ensuring that all tokens have at least one token of look-ahead. MLA size $M=1$.}
    \label{fig:mla}
\end{figure}

\subsection{Minimum look-ahead}
The chunk-aware streaming achieves a good balance between look-ahead and latency, and has demonstrated excellent performance in streaming ASR tasks~\cite{noroozi2024stateful}. However, through preliminary experiments, we found a problem when applied to the G2PnP task. That is, since the last tokens of each chunk have no look-ahead, the tokens are prone to generating outputs that are inconsistent with the subsequent tokens. To tackle this issue, we introduce a simple, but effective method named \textit{minimum look-ahead} (MLA). The concept of MLA is shown in Fig.~\ref{fig:mla}. As shown in the figure, MLA allows a token to refer to $M$ future tokens outside the current chunk. As a result, even the last token in a chunk can see the future tokens, which we expect to lead to consistent prediction at chunk boundaries. \sedit{In comparison with standard chunk-aware streaming, the look-ahead size for all tokens is increased by $M$, resulting in a look-ahead size ranging from $C+M-1$ to $M$.}  
\sedit{Note that we apply this method only to the first self-attention layer.} This is because, from the second layer onward, future tokens acquire additional future dependencies through the self-attention of the preceding layers, whereas future tokens in the first layer remain free from such dependencies.

\subsection{Inference}
In the inference stage, the model takes a stream of grapheme tokens as input and begins chunk-level prediction of phoneme and prosodic labels as soon as the first $C$ grapheme tokens become available. When MLA is enabled, the model waits $C+M$ grapheme tokens in total. 

\section{Experiments}
\label{sec:exp}

\subsection{Experimental conditions}
\subsubsection{Data preparation}
To confirm the effectiveness of the proposed method, we conducted experiments of the G2PnP task. Since the advantage of the proposed method over~\cite{dekel2024speak} lies in its applicability to unsegmented languages, we conducted experiments on Japanese, which is one such language. Specifically, we used the transcriptions in the ReazonSpeech dataset~\cite{ReazonSpeech} for the training of the proposed models. The number of the transcriptions for training and validation was 14,960,911 and 4,802, respectively. To prepare the target PnP sequence for training, we used a \yedit{MeCab}-based non-streaming morphological analysis model~\cite{mecab}, followed by a DNN-based prosodic label prediction model~\cite{park2022unified}. More precisely, the former \sedit{first performs text normalization and then} predicts word-level features such as surface, reading, \yedit{pronunciation}, and part-of-speech using the Japanese dictionary Unidic~\cite{unidic}. The latter predicts intonation phrase boundaries (IPs, ``\#''), accent phrase boundaries (APs, ``/''), and accent nucleus (ANs, ``*'') from these features. The \sedit{prosodic label prediction model} is trained on 80,061 manually-annotated prosodic labels. Further details can be found in ~\cite{park2022unified}. We refer to this G2PnP model as \textbf{Dict-DNN}. 

For evaluation, it is desirable to use a dataset in which high-quality phoneme and prosodic labels are provided for graphemes across a variety of domains. To this end, we created expert annotations of phonemes and prosodic labels for a total of 2,722 sentences drawn from six domains—chat, interview, news, novel, practical book, and SNS—and used this dataset for evaluation. We refer to this dataset as \textbf{6D-Eval}.

\subsubsection{Model details}
We configured the streaming Conformer with eight layers and a hidden dimension of 512. To facilitate integration with an LLM, the tokenizer employed was the Byte Pair Encoding (BPE) tokenizer bundled with the CALM2-7B-Chat model, publicly released on Hugging Face\footnote{\url{https://huggingface.co/cyberagent/calm2-7b-chat}}. For the up-sampling of token IDs, we employed a simple repetition-based method with an up-sampling factor of eight. The up-sampling factor was determined based on preliminary experiments\sssedit{, motivated by the observation that smaller factors were insufficient to cover all PnP tokens corresponding to a single grapheme token}. During training, we employed dynamic batching such that each mini-batch contained up to 8,192 tokens. 
The learning rate was set to 1e-4 and was exponentially decayed to 1e-5 over 1.2M steps. \sedit{Intermediate CTC losses were applied at the 2nd, 4th, and 6th layers, with their weights set to one third of the final CTC loss.}

For the comparison of streaming G2PnP models, we used a streaming version of Dict-DNN models as the baseline. Since the Dict-DNN model involves non-streamable processes such as morphological analysis, we constructed streaming versions of these baseline models by segmenting the input grapheme sequence into fixed-size chunks and sequentially processing them with the non-streaming Dict-DNN model. For the streaming Dict-DNN models, we used chunk sizes of 5, 10, and 20 \sedit{tokens}. For the proposed \textbf{CC-G2PnP}, we trained six streaming models in total, using chunk sizes of 2 and 5 tokens and minimum look-ahead sizes of 0, 1, and 2 tokens. The past context size was set to 10 tokens for all the proposed models. In addition to the streaming models, we also trained non-streaming versions of Dict-DNN and CC-G2PnP for reference.

\begin{table*}[htbp]
\centering
\scalebox{0.93}{
\begin{tabular}{llccccccc}
\toprule
\multirow{2}{*}{Type} & \multirow{2}{*}{Model} 
 & \multicolumn{3}{c}{Config} & \multicolumn{3}{c}{CER (SER)} & CPU processing time [\si{\second}] \\
\cmidrule(lr){3-5} \cmidrule(lr){6-8} \cmidrule(lr){9-9}
 &  & $P$ & $C$ & $M$ & PnP & Norm. PnP & Phoneme & Start \\
\midrule
\multirow{9}{*}{Streaming}
 & Dict-DNN-5
   & 0 & 5  & 0 & 6.67 (84.7) & 5.90 (82.8) & 1.54 (22.4) & 5$\tau$ + 0.0331 \\
 & Dict-DNN-10 & 0 & 10 & 0 & 3.58 (62.2) & 2.97 (57.2) & 0.86 (13.0) & 10$\tau$ + 0.0334  \\
 & Dict-DNN-20 & 0 & 20 & 0 & 2.28 (45.9) & 1.72 (36.0) & 0.56 (8.7)  & 20$\tau$ + 0.0352  \\
\cmidrule(lr){2-9}
 & CC-G2PnP-2-0
   & 10 & 2  & 0 & 2.44 (51.6) & 1.77 (39.7) & 0.82 (16.2) & 2$\tau$ + 0.0375  \\
 & CC-G2PnP-2-1 & 10 & 2  & 1 & 1.90 (43.4) & 1.36 (30.0) & 0.55 (9.0)  & 3$\tau$ + 0.0395  \\
 & CC-G2PnP-2-2 & 10 & 2  & 2 & 1.85 (42.4) & 1.33 (29.8) & 0.53 (8.7)  & 4$\tau$ + 0.0395  \\
 & CC-G2PnP-5-0 & 10 & 5  & 0 & 2.01 (45.3) & 1.45 (33.4) & 0.62 (11.3) & 5$\tau$ + 0.0395  \\
 & CC-G2PnP-5-1 & 10 & 5  & 1 & \textbf{1.79} \textbf{(41.4)} & \textbf{1.28} \textbf{(28.6)} & \textbf{0.52} \textbf{(8.4)}  & 6$\tau$ + 0.0399  \\
 & CC-G2PnP-5-2 & 10 & 5  & 2 & 1.80 \textbf{(41.4)} & 1.30 (28.9) & 0.53 (8.6)  & 7$\tau$ + 0.0399  \\
\midrule
\multirow{2}{*}{\textcolor{gray}{Non-streaming}}
 & \textcolor{gray}{Dict-DNN-NS} & \multicolumn{3}{c}{\textcolor{gray}{$\infty$}} 
   & \textcolor{gray}{1.71 (40.4)} & \textcolor{gray}{1.18 (26.4)} & \textcolor{gray}{0.40 (6.4)}  
   & \textcolor{gray}{$N\tau$ + 0.0327}  \\
 & \textcolor{gray}{CC-G2PnP-NS} & \multicolumn{3}{c}{\textcolor{gray}{$\infty$}} 
   & \textcolor{gray}{1.80 (42.0)} & \textcolor{gray}{1.33 (29.9)} & \textcolor{gray}{0.48 (7.6)}  
   & \textcolor{gray}{$N\tau$ + 0.0711}\\
\bottomrule 
\end{tabular}
}
\caption{Comparison of CER, SER, and computational time. 
$P$, $C$, and $M$ denote past context size, chunk size, and minimum look-ahead size, respectively. 
(Norm.) PnP denotes (normalized) phoneme and prosodic labels. \textit{Start} indicates the latency until the first output is obtained when combined with the LLM, and $\tau$ [\si{\second}] denotes the time for the LLM to generate one token. $N$ denotes the number of tokens in one sentence. \textbf{Bold} font means the best score among streaming models.}
\label{tab:objective}
\end{table*}

\begin{table}[htbp]
\vspace{-1mm}
\centering
\scalebox{0.93}{
\begin{tabular}{llc}
\toprule
Type & Model & MOS \\
\midrule
\multirow{4}{*}{Streaming}
& Dict-DNN-5  & 2.73 $\pm$ 0.11 \\
& Dict-DNN-10 & 3.35 $\pm$ 0.11 \\
& CC-G2PnP-5-0  & 3.81 $\pm$ 0.10 \\
& CC-G2PnP-5-1  & \textbf{4.02 $\pm$ 0.09} \\
\midrule
\multirow{2}{*}{Non-streaming}
& Dict-DNN-NS           & 4.07 $\pm$ 0.09 \\
& CC-G2PnP-NS      & 4.02 $\pm$ 0.09 \\
\midrule
- & GT-Label      & 4.16 $\pm$ 0.07 \\
\bottomrule
\end{tabular}
}
\caption{Speech naturalness MOS with 95\% confidence intervals. \textbf{Bold} is the best score among streaming models.}
\label{tab:mos}
\vspace{-3mm}
\end{table}

\vspace{-1mm}
\subsection{G2PnP accuracy}
To objectively evaluate the performance of the G2PnP task, we used the CER and SER between the ground-truth and predicted PnP sequences. To independently evaluate phoneme prediction, we also computed CER and SER on sequences with the prosodic labels removed. Furthermore, since the estimation of IP and AP is highly speaker-dependent and strict ground-truth labels are difficult to define, we also computed scores under the condition that IP and AP were treated as identical, referred to as \textit{Norm. PnP}. 

The results are shown in the CER (SER) section of Table \ref{tab:objective}. As shown in the table, the proposed streaming G2PnP models with positive MLA significantly outperformed the baseline models and those without MLA in all metrics. When comparing models with and without MLA, we can see that, for both chunk sizes of 2 and 5, the adoption of MLA leads to improvements across all metrics. Among the proposed methods, the models with a chunk size of 5 and an MLA of 1 or 2 achieved the best performance, yielding scores that are close to the non-streaming Dict-DNN model. 
These results confirm the effectiveness of the proposed methods, including MLA for streaming G2PnP. 

\subsection{Processing time}
From a streaming perspective, processing speed is also an important evaluation metric. Therefore, for each model, we measured the average time until the first token was generated (\textit{Start}). 
All measurements were conducted on an Apple M4 Pro CPU. The results are shown in \ssedit{the CPU processing time section of} Table~\ref{tab:objective}. Comparing the streaming Dict-DNN and proposed models, the time required for generating a single chunk by the G2PnP model alone is of the same order (0.03--0.04 [\si{\second}]). When considering integration with LLM, the proposed method attains its best performance with a latency of just 6 tokens (chunk size of 5 and MLA size of 1), whereas the conventional method fails to match this performance even at 20 tokens. These results suggest that the proposed method enables reduced latency when combined with an LLM, while still achieving substantially better performance than the conventional approach.

\subsection{TTS perceptual quality}
The ultimate goal of streaming G2PnP is not merely to improve CER or SER of the generated PnP sequence, but to generate speech that sounds natural to humans when combined with a downstream TTS system. Therefore, we conducted a subjective listening test to evaluate the naturalness of the speech when the proposed method was combined with TTS.
The TTS model was based on NANSY-TTS~\cite{choi2022nansy++}, modified to take phoneme and prosodic label sequences as text input. To ensure that the generated speech faithfully reflected the input phoneme and prosodic label sequences, the TTS model was trained on an internally constructed Japanese speech corpus consisting of 173,987 samples \sedit{with manually-annotated phonemic and prosodic labels}, totaling 207.96 hours. Although the training data included 17 speakers, the evaluation was conducted using the speech of a single female speaker in order to eliminate speaker-related effects. Note that a streaming TTS model was not employed, since the purpose of this experiment was to assess the quality of pronunciation and prosody in speech synthesized from the PnP sequences generated by the proposed methods, rather than to evaluate latency.

The subjective listening test was conducted using the mean opinion score (MOS) on a five-point scale: 1 = Bad (reading and prosody are highly unnatural); 2 = Poor; 3 = Fair; 4 = Good; 5 = Excellent (reading and prosody sound natural, as in human speech). Native Japanese listeners were recruited to assess the speech quality in terms of prosodic naturalness and reading accuracy. During the listening test, the raters were also presented with the grapheme text to facilitate more accurate judgments. In total, 15 raters participated, and 50 sentences were randomly selected from \textbf{6D-Eval}, \ssedit{uniformly sampled across its six domains.} \sedit{Considering the evaluation cost}, we compared PnP sequences generated by four streaming models \sedit{with chunk sizes of 5 and 10}, two non-streaming models, and the ground-truth \ssedit{label (GT-Label)} sequence. 

Table \ref{tab:mos} presents the MOS with 95\% confidence intervals. The results indicate that the proposed streaming models (CC-G2PnP) substantially outperform the streaming baselines (Dict-DNN) in terms of reading and prosody naturalness. In particular, CC-G2PnP-5-1 achieved the highest score among streaming models (4.02), which is very close to the non-streaming counterparts (Dict-DNN-NS: 4.07, CC-G2PnP-NS: 4.02) and the ground-truth labels (4.16). These findings suggest that the proposed method can generate PnP sequences with reading and prosody naturalness comparable to that of non-streaming systems, and the trend is consistent with that observed in G2PnP accuracy.

\begin{table}[t]
\vspace{-1mm}
\centering
\scalebox{0.93}{
\begin{tabular}{lcc}
\toprule
\multirow{2}{*}{Data size} & \multicolumn{2}{c}{CER (SER)} \\ 
\cmidrule(lr){2-3}
 & PnP & Phoneme \\
\midrule
1\%   & 4.55 (64.0) & 1.97 (24.8) \\
10\%  & 2.34 (47.8) & 0.71 (10.6) \\
100\% & \textbf{1.79 (41.4)} & \textbf{0.52 (8.4)} \\
\bottomrule
\end{tabular}
}
\caption{Impact of training data size on the accuracy of G2PnP. CC-G2PnP-5-1 was used for all the evaluations.}
\label{tab:data}
\vspace{-3mm}
\end{table}

\subsection{Impact of training data size}
Since the proposed method learns G2PnP in a data-driven manner, examining the relationship between performance and the amount of training data is important. To this end, we evaluated the model using 10\% (1,496,091 samples) and 1\% (149,609 samples) of the training data. All other settings, including the number of training steps, were kept unchanged. We used CC-G2PnP-5-1 for all the evaluations. The results are shown in Table~\ref{tab:data}. We can see that the performance improves consistently as the training data increases, with substantial gains observed when moving from 1\%  to larger portions of the dataset. These findings suggest that the proposed method can effectively leverage larger training data to achieve higher performance.

\section{Conclusions}
\label{sec:majhead}
We proposed CC-G2PnP, a streaming model that converts graphemes into phonemic and prosodic labels. The model is based on the Conformer-CTC architecture, and is applicable to unsegmented languages. Experiments showed the effectiveness of the proposed methods in the streaming \yedit{G2PnP} task. Since the proposed method does not rely on external dictionaries, it requires a large amount of training data to cover a diverse grapheme vocabulary. Therefore, an important direction for future work is to effectively integrate external knowledge from dictionaries or LLMs.

\vfill\pagebreak

{\small
\bibliographystyle{IEEEbib}
\bibliography{refs}
}

\end{document}